\documentclass[letterpaper]{article} 
\usepackage{aaai2026}  
\usepackage{times}  
\usepackage{helvet}  
\usepackage{courier}  
\usepackage[hyphens]{url}  
\usepackage{graphicx} 
\usepackage{natbib}  
\usepackage{caption} 
\usepackage{algorithm}
\usepackage{algorithmic}
\usepackage{amsmath}

\usepackage{newfloat}
\usepackage{listings}
\DeclareCaptionStyle{ruled}{labelfont=normalfont,labelsep=colon,strut=off} 
\floatstyle{ruled}
\newfloat{listing}{tb}{lst}{}
\floatname{listing}{Listing}
\title{Backdoor Attacks and Defenses in Computer Vision Domain: A Survey}
\author{
    Bilal Hussain Abbasi, Yanjun Zhang, Leo Zhang, Shang Gao\\
}

\begin{document}

\maketitle

\begin{abstract}
Backdoor (trojan) attacks embed hidden, controllable behaviors into machine-learning models so that models behave normally on benign inputs but produce attacker-chosen outputs when a trigger is present. This survey reviews the rapidly growing literature on backdoor attacks and defenses in the computer-vision domain. We introduce a multi-dimensional taxonomy that organizes attacks and defenses by injection stage (dataset poisoning, model/parameter modification, inference-time injection), trigger type (patch, blended/frequency, semantic, transformation), labeling strategy (dirty-label vs. clean-label / feature-collision), representation stage (instance-specific, manifold/class-level, neuron/parameter hijacking, distributed encodings), and target task (classification, detection, segmentation, video, multimodal). For each axis we summarize representative methods, highlight evaluation practices, and discuss where defenses succeed or fail. For example, many classical sanitization and reverse-engineering tools are effective against reusable patch attacks but struggle with input-aware, sample-specific, or parameter-space backdoors and with transfer via compromised pre-trained encoders or hardware bit-flips. We synthesize trends, identify persistent gaps (supply-chain and hardware threats, certifiable defenses, cross-task benchmarks), and propose practical guidelines for threat-aware evaluation and layered defenses. This survey aims to orient researchers and practitioners to the current threat landscape and pressing research directions in secure computer vision.
\end{abstract}


\section{Introduction}
The security of deep neural networks (DNNs) is a pressing practical concern as AI systems, increasingly driven by machine learning and deep learning are deployed across safety-critical and privacy-sensitive domains. In practice, development and deployment efforts often prioritize functionality and performance, while adversarial risk receives less operational attention; this gap leaves systems exposed to a range of attacks that can undermine reliability and trust. Improving DNN security is therefore essential to ensure these systems behave safely and predictably in real-world settings \cite{szegedy2013intriguing,goodfellow2014explaining,amodei2016concrete,gu2019badnets,liu2017trojaning,wang2019neural}.

Threats to DNNs take many forms. A well-studied example in computer vision is adversarial examples, where small, often imperceptible perturbations to an input cause gross model misbehavior; such perturbations can be produced automatically and are typically invisible to human observers \cite{goodfellow2014explaining}. A related but distinct class of threats are backdoor (or trojan) attacks: here the adversary implants a hidden association into the model during training so that, while the model performs normally on benign inputs, the presence of a specific trigger (a visible patch, a subtle pixel perturbation, a semantic attribute, etc.) causes a controlled, attacker-chosen output. Backdoors are especially insidious because they remain dormant until the trigger appears, and they can be injected via poisoned training data, compromised pre-trained encoders, or maliciously altered model checkpoints \cite{gu2019badnets,gao2020backdoor}.

Backdoor attacks impact a wide range of computer-vision tasks and the way they manifest depends heavily on the task. In image classification, the backdoor typically maps any input containing the trigger to a target label (e.g., every image with a small symbol or patch is classified as ``stop'') \cite{gu2019badnets,chen2017targeted,liu2017trojaning}, enabling trivial but dangerous bypasses of security checks. In object detection and semantic segmentation the attack surface is richer: attackers may aim to hide objects (cloaking) \cite{ma2022dangerous}, cause targeted mislocalization or mislabeling of regions \cite{lan2024influencer,li2021hidden}, or selectively alter instance-level predictions while leaving other outputs intact \cite{chan2022baddet}. 

These dense and structured prediction tasks introduce additional challenges for both attackers and defenders because the model must reason about spatial context, multi-scale features, and pixel or proposal-level consistency. Consequently, defenses that work well for whole-image classification do not automatically translate to detectors or segmenters, motivating task-specific analyses and tailored mitigation strategies.

This survey adopts a mirrored, multi-axis taxonomy that organizes both attacks and defenses along the same conceptual dimensions: injection stage, trigger type, labeling strategy, representation stage, and target task. Doing so clarifies which defenses plausibly counter which attacks, and where gaps remain. For example, dataset sanitization and spectral/activation filters are effective defenses for many dirty-label and patch-style training poisons, while parameter-space or supply-chain attacks demand provenance, checkpoint auditing, and hardware integrity measures. Inference-time defenses (perturbation-based detectors, input purification, transformation consistency checks) are attractive for deployment but trade latency and false positives; they also can be circumvented by input-aware or transformation-robust triggers.

We focus on backdoor threats and countermeasures specifically in the computer-vision ecosystem: covering image classification, object detection, semantic segmentation, video, and related dense-prediction tasks. Our survey (1) synthesizes representative attacks and defenses under a unified taxonomy; (2) summarizes evaluation protocols and common metrics; (3) identifies practical gaps (sample-specific triggers, parameter/hardware attacks, transfer via pre-trained encoders, lack of standardized cross-task benchmarks); and (4) offers concrete recommendations for layered defenses and more rigorous, attacker-aware evaluation. Throughout we highlight works that demonstrate practical physical robustness, transferability, or supply-chain realism, and we emphasize where defensive claims should be evaluated under adaptive attacks.

After this introduction, we present (i) a preliminaries section that fixes notation, threat models and metrics for evaluation,(ii) an organized review of attack families (Section: Backdoor Attacks), (iii) defenses grouped by the mirrored taxonomy (Section: Backdoor Defenses), and (iv) a conclusion section that includes discussion, open problems and recommendations. 

\section{Preliminaries}
\label{sec:preliminaries}

This section fixes notation, standard threat models, and common evaluation metrics used throughout the survey.

\subsection{Notation}
Let \(\mathcal{D}=\{(x_i,y_i)\}_{i=1}^N\) denote a supervised training dataset of input--label pairs, where \(x\in\mathcal{X}\) (images) and \(y\in\mathcal{Y}\) (discrete class labels or structured outputs for detection/segmentation).  
A model is denoted \(f_\theta:\mathcal{X}\to\mathcal{Y}\) and is parameterized by weights \(\theta\). For detectors and segmenters the output space \(\mathcal{Y}\) represents structured predictions (e.g., bounding boxes and class scores, or pixel-wise labels).

We write a trigger application operator as \(T(\cdot;\tau)\), where \(\tau\) parameterizes the trigger (e.g., patch location and content, blending strength, frequency coefficients, or parameters of an input-aware generator). A poisoned training set \(\mathcal{D}_p\) is obtained by inserting poisoned examples
\[
\{(T(x_i;\tau_i), y'_i)\}_{i\in\mathcal{I}_p}
\]
into \(\mathcal{D}\). The poisoning rate is
\[
\rho \;=\; \frac{|\mathcal{I}_p|}{N},
\]
and \(y'_i\) may equal the original label (``clean-label'') or be replaced by an attacker-chosen target label (``dirty-label'').

\subsection{Threat-model axes}
Backdoor threats are best described along multiple, often orthogonal, axes. We use the following taxonomy when describing attacks and defenses.

\begin{enumerate}
  \item \textbf{Injection stage:}
    \begin{itemize}
      \item \emph{Training/data poisoning:} adversary injects poisoned samples into \(\mathcal{D}\).
      \item \emph{Model/parameter modification:} adversary tampers with model weights or checkpoints (supply-chain compromise, targeted weight perturbations, bit-flips).
      \item \emph{Inference-time injection:} adversary applies triggers only at inference (input-aware or dynamic triggers).
    \end{itemize}

  \item \textbf{Trigger type:}
    \begin{itemize}
      \item \emph{Patch / sticker:} localized, often visibly placed pattern.
      \item \emph{Blended / invisible / frequency:} diffuse or spectral-domain encodings.
      \item \emph{Semantic:} natural attributes/accessories (e.g., glasses, stickers with plausible semantics).
      \item \emph{Transformation-based:} geometric warps or other transformation triggers.
    \end{itemize}

  \item \textbf{Labeling strategy:}
    \begin{itemize}
      \item \emph{Dirty-label:} poisoned inputs are mislabeled to the attack target.
      \item \emph{Clean-label / feature-collision:} poisoned inputs retain plausible labels; the attack manipulates internal representations.
    \end{itemize}

  \item \textbf{Representation stage:}
    \begin{itemize}
      \item \emph{Instance-specific / input-aware:} trigger depends on input content.
      \item \emph{Class-level / manifold:} attack collapses representations of many inputs toward the target manifold.
      \item \emph{Neuron / parameter hijacking:} direct manipulation of weights or subnetworks to embed triggers.
      \item \emph{Distributed encodings:} trigger effects distributed across layers/filters.
    \end{itemize}

  \item \textbf{Defender access assumptions:}
    \begin{itemize}
      \item \emph{White-box:} defender has model internals and/or training data.
      \item \emph{Black-box / query-only:} defender only queries the model.
      \item \emph{Data-available vs.\ model-only:} whether defenders can inspect training data (important for sanitization).
    \end{itemize}
\end{enumerate}

\subsection{Evaluation metrics}
We summarize the most common metrics used to evaluate attacks and defenses.

\paragraph{Clean performance.} Clean accuracy (for classification) or standard utility metrics (e.g., mAP for detection, mean IoU for segmentation) measured on a held-out benign test set:
\[
\text{CA} \;=\; \Pr_{(x,y)\sim\mathcal{D}_\text{test}}\big[f_\theta(x)=y\big].
\]

\paragraph{Attack success rate.} For a \emph{targeted} backdoor (target label \(y_\text{tgt}\)), the attack success rate (ASR) is defined as the probability that a trigger-bearing input is mapped to the attacker target:
\[
\text{ASR} \;=\; \Pr_{x\sim\mathcal{D}_\text{test}}\big[f_\theta(T(x;\tau))=y_\text{tgt}\big].
\]
For detection/segmentation attacks this generalizes to the fraction of triggered inputs where the desired misbehavior (e.g., missed detection, wrong mask) occurs.

\paragraph{Utility-stealth tradeoff.} Report the change in clean performance caused by the attack (or by a defense):
\[
\Delta\text{CA} \;=\; \text{CA}_\text{post} - \text{CA}_\text{pre}.
\]
Smaller degradation (in absolute value) indicates higher stealth for the attacker or lower cost for the defender.

\paragraph{Operational metrics for defenses.} For detection or mitigation mechanisms report standard operating metrics (True Positive Rate / False Positive Rate for trigger detection), computational/latency overhead, and any degradation to clean inputs introduced by runtime purifiers or transformations.

\subsection{Defender goals and constraints}
Typical defender objectives include:
\begin{itemize}
  \item \textbf{Detection:} decide whether a delivered dataset or checkpoint contains a backdoor.
  \item \textbf{Localization / reverse-engineering:} estimate the trigger or its support region.
  \item \textbf{Repair:} remove or mitigate the backdoor while retaining utility (e.g., fine-tuning, pruning, data-sanitization, or model surgery).
  \item \textbf{Certification:} obtain provable guarantees for a restricted family of triggers (e.g., local patch certificates).
\end{itemize}

Practical constraints that shape defense design are limited access to a trusted clean holdout, compute budget, the potential for adaptive attackers, and the need for cross-task robustness (defenses effective across classification, detection and segmentation).

\section{Backdoor Attacks}
\label{sec:backdoor-attacks}

This section organizes existing backdoor attacks in computer vision using a concise, multi-dimensional taxonomy. The objectives are twofold: (1) To characterize the design choices and threat models that distinguish different attacks, and (2) To provide a framework that directly relates attack types to potential defense strategies in the following section. Because backdoor research spans multiple aspects, we consider several independent dimensions rather than a single linear ordering; a given work may therefore be categorized under multiple headings depending on its characteristics.

The dimensions used throughout this chapter are:
\begin{itemize}
\item \textbf{Injection stage:} The point at which the adversary implants the backdoor, such as poisoning the training dataset, modifying model parameters post-training, or manipulating inputs at inference.
\item \textbf{Trigger type:} The form of the backdoor trigger, which can be a visible patch or sticker, blended pattern or noise, a semantic attribute, or an input transformation.
\item \textbf{Labeling strategy:} Whether poisoned samples are relabeled (dirty-label) or retain their correct labels (clean-label / feature-collision).
\item \textbf{Representation stage:} How the backdoor is encoded within the model, for example via instance-specific representations, class-wide shifts in feature space, neuron/parameter hijacking, or distributed multi-layer encodings.
\item \textbf{Target task:} The vision task affected by the attack, including image classification, object detection, semantic segmentation, video, or other dense-prediction tasks.
\end{itemize}

Applying this taxonomy enables a systematic comparison of attacks, helps to identify common assumptions and limitations, and allows for a critical evaluation of whether existing defenses address the full threat landscape. Where relevant, we also highlight attacks that are particularly stealthy, transferable, or effective in physical settings, and note areas where multiple attack families overlap.

\subsection{Injection Stage}
The injection stage refers to the point in the machine learning pipeline at which a backdoor is embedded. This dimension is crucial because it determines the attacker’s operational requirements, potential stealth, and the corresponding defense strategies. Existing works can be broadly grouped into dataset poisoning, model parameter modification, and on-the-fly injection at inference.

\subsubsection{Dataset Poisoning (training phase)}  
Dataset poisoning which comprises of the modification or injection of a small fraction of training samples, so the learned model encodes a trigger to target mapping remains the canonical and most intensively studied backdoor injection stage in computer vision. The seminal BadNets work \cite{gu2019badnets} demonstrated that simply stamping a visible patch onto a few training images and relabelling them can produce near-perfect attack success rates while leaving benign accuracy essentially unchanged. 

Follow-up works generalized this basic method in multiple directions: Chen et al. \cite{chen2017targeted} formalized targeted data-poisoning strategies and proposed blending-style triggers that reduce perceptual conspicuity; Turner et al. \cite{turner2019label} showed how to construct \emph{label-consistent} (clean-label) poisons that keep annotations intact by using adversarial perturbations or generative synthesis, dramatically increasing stealth; and Liu et al. \cite{liu2020reflection} introduced \emph{natural} triggers (e.g., reflections) that exploit real-world image phenomena to hide backdoors in realistic settings. 

More recent training-phase attacks have emphasised trigger diversity and stealth: Input-Aware Dynamic attacks \cite{nguyen2020inputaware} train a generator to produce per-sample, input-conditioned triggers that defeat defenses assuming a single universal trigger, while WaNet \cite{nguyen2021wanet} uses imperceptible, warping-based transformations to craft triggers that are both visually benign and hard to detect by inspection or many automated detectors. Historical and complementary lines of work, for example, Trojaning attacks \cite{liu2017trojaning}) further show that poisoning or maliciously curated training pipelines can implant trojans via subtle data/model changes that survive fine-tuning and transfer learning. These works illustrate the rich design space of training-phase poisoning (visible or invisible, label-corrupting or clean-label, sample-agnostic or input-specific) and explain why dataset hygiene and pre-training sanitisation remain foundational defense requirements.

\subsubsection{Model Parameter Modification (post-training)}  
Model-parameter modification attacks bypass data-level sanitation by tampering directly with model weights or delivered checkpoints, turning the model artifact itself into the attack vector. Early demonstrations showed that adversaries who can edit a released checkpoint can implant stealthy trojans that map trigger-bearing inputs to attacker-chosen outputs while preserving benign accuracy; this class of threat has since diversified into several distinct variants. In federated and collaborative learning, model-replacement and model-poisoning attacks enable a single malicious client (or a small colluding set) to push a poisoned global model via crafted parameter updates, often evading naive aggregation and outperforming classic data poisons \cite{bagdasaryan2020backdoor}. 

Distributed strategies, exemplified by the DBA attack of Xie et al. \cite{xie2019dba}, show that distributed backdoor patterns can be injected across multiple clients, thereby reducing visibility at the individual level. Supply-chain threats exploit the prevalence of pre-trained checkpoints: attackers can publish or compromise foundation encoders so that downstream fine-tuned models inherit the backdoor across tasks, as demonstrated by BadEncoder \cite{jia2022badencoder} and subsequent works showing that backdoored pre-trained models transfer trojans broadly to downstream classifiers \cite{shen2021backdoor}. 

More surgical weight-poisoning attacks craft layerwise parameter perturbations to plant deeper, harder-to-erase backdoors in released checkpoints \cite{li2021backdoor}. These parameter-space attacks expand the adversary’s surface from collaborative training and model repositories to device memory, and they pose distinctive challenges for detection and mitigation since they may leave no clear signature in the training data and can endure through fine-tuning or model compression \cite{wei2023aliasing}. Defending against them therefore requires provenance controls, parameter-space auditing, and integrity checks in addition to dataset hygiene.

A complementary line of work targets deployment- or hardware-level manipulation rather than the model repository itself. Training-assisted bit-flip attacks, introduced by Dong et al. \cite{dong2023one}, craft so-called `high-risk’ models which appear benign but are primed such that a single bit flip during deployment activates a backdoor. Similarly, Cai et al. propose WBP \cite{cai2024wbp}, a hardware-based weight-poisoning method that leverages Rowhammer-triggered bit flips during fine-tuning to inject stealthy, dense backdoors. Other efforts, such as ONEFLIP by Li et al. \cite{li2025oneflip}, show that even a single bit alteration at inference can suffice to implant a robust Trojan into full-precision models. 

Such threats are particularly insidious because they bypass data-centric defenses like dataset sanitization and often survive naive fine-tuning. Consequently, defending against them requires mechanisms beyond input inspection, specifically, checkpoint provenance, parameter-space anomaly detection, robust aggregation in federated settings, and hardware-level integrity safeguards.
These works highlight that post-training parameter modification constitutes a versatile, high-impact threat vector, one that necessitates both supply-chain and deployment-level security measures in computer vision pipelines.


\subsubsection{On-the-fly Injection (inference phase)}  
On-the-fly (inference-time) backdoor attacks broaden the adversary’s capabilities by placing the trigger or a necessary perturbation, only at test time, rather than relying on a single, fixed pattern learned during training. A key subclass are input-conditioned or dynamic-trigger attacks that produce a bespoke trigger per input; the Input-Aware Dynamic Backdoor Attack \cite{nguyen2020inputaware} trains a generator that produces instance-specific triggers conditioned on each image, making triggers non-reusable and substantially degrading the effectiveness of reverse-engineering or fixed-trigger detectors. 

Another important direction fuses adversarial perturbations with trojans, requiring both a poisoned model and a carefully crafted test-time perturbation to activate the backdoor. Synergistic approaches such as AdvTrojan \cite{9671964} show that this coupling can bypass defenses which treat adversarial noise and trojans as separate threats. More recently, researchers have shown that test-time injection can be applied in broader settings, for example, universal image perturbations that trigger behaviours in multimodal models, such as AnyDoor \cite{lu2024test} or more general test-time injection schemes that do not require prior access to model training.

These inference-phase attacks have spurred a parallel literature on test-time detection and mitigation. Practical detectors operating at inference inspect robustness and prediction consistency under controlled corruptions (TeCo), perform in-flight reverse-engineering and source-class inference on suspicious inputs, or adopt unified statistical frameworks with provable false-positive guarantees for detecting anomalous inputs at query time \cite{liu2023detecting,li2022test,xian2023unified}. 

Other work studies test-time repair strategies that both detect and then locally neutralise backdoor activations (e.g., detection-plus-pruning pipelines) using only a small batch of available test inputs \cite{guan2024backdoor}. The inference-time threat model therefore significantly widens the attacker’s practical options and implies that robust evaluation of backdoor defenses must include online scenarios (dynamic triggers, adversarial+trojan hybrids, and universal test-time perturbations) as well as defenses tailored for low-access, black-box query settings.

\subsection{Trigger Types}
The trigger type refers to the specific visual or structural pattern embedded into the input to activate the backdoor. The nature of the trigger directly impacts an attack’s stealthiness, generalizability, and resistance to defenses. Triggers in computer vision backdoor attacks can be broadly categorized into patch-based, blended, semantic, and input transformation–based designs.

\subsubsection{Patch-based Triggers}
Patch-based triggers are among the most intuitive and widely studied backdoor mechanisms in computer vision: an attacker embeds a compact, often visible pattern (a patch, sticker, or printed object) into a subset of training images and typically relabels them so that the model learns a strong association between the patch and the adversary's chosen target. The canonical demonstration is BadNets \cite{gu2019badnets}, which showed that stamping a small, consistently placed square into a few training images can produce very high attack success rates while leaving clean accuracy largely intact. Follow-up work extended the patch paradigm to more realistic and challenging settings: Chen et al. \cite{chen2017targeted} formalized targeted data-poisoning attacks that can be implemented in the physical world (e.g., eyeglass-like frames for face recognition) and introduced blending/placement choices to improve stealth. 

Empirical studies then systematically evaluated the feasibility of physical-object triggers (sunglasses, stickers, T-shirts, Post-it notes), showing that real-world triggers can be highly effective under many conditions but are sensitive to placement, viewpoint and imaging quality \cite{wenger2021backdoor,li2021backdoor}. For dense prediction tasks, several works adapted patch-style backdoors to object detectors, for example, BadDet \cite{chan2022baddet} and related studies show that localized stickers or patches can induce targeted misdetections or cloaking effects in popular detectors \cite{ma2022dangerous}. 

More recent attacks have improved physical robustness by using variable-size triggers, adversarial training with physical noise, or specially-crafted trigger collections to tolerate scale, rotation and lighting variation \cite{qian2023robustbackdoorattacksobject,li2021backdoor}. Research efforts have also produced publicly usable ``natural backdoor'' datasets and identification pipelines that locate naturally co-occurring physical objects in large image corpora, which both facilitate and expose the threat of physical patch triggers \cite{wenger2022natural}. 

Hence, this body of work shows that patch-based backdoors combine high effectiveness with simple deployability in many scenarios, properties that continue to make them a major practical security concern for deployed vision systems.

\subsubsection{Blended Triggers}  
Blended triggers hide the adversary’s signal by smoothly integrating the trigger into the host image so that poisoned samples remain visually similar to their clean counterparts while still producing a deterministically malicious behavior at inference. Early work formalized blending as a practical way to reduce conspicuity (e.g., low-opacity overlays and mixed-image blends) and demonstrated that such triggers can be learned reliably from a small poisoning rate \cite{chen2017targeted}. 

Subsequent advances pushed stealth further along several axes. Label-consistent \cite{turner2019label} and hidden-trigger attacks \cite{saha2020hidden} showed that triggers can be embedded without visibly altering the semantic label of poisoned samples, making human inspection and simple dataset filtering ineffective. Other attacks make triggers instance-specific, i.e., generating a unique, imperceptible perturbation per poisoned sample via image steganography or learned generators, which defeats defenses that assume a single universal pattern \cite{li2021invisible,wang2024ghostencoder}. 

A complementary line of work investigates the frequency domain, showing that many pixel-space triggers leave characteristic high-frequency artifacts, and proposing frequency-aware designs that smooth or distribute spectral energy across bands to evade detection \cite{zeng2021rethinking,wang2021backdoor}. More recent work integrates these ideas into fully dynamic, invisible triggers (and sparse+invisible constructions) that are both label-consistent and robust to routine augmentations, demonstrating high attack success while substantially narrowing the defensive window for detection and reverse-engineering \cite{narisada2023fully,gao2024backdoor}.

Blended-trigger research highlights a persistent trade-off: improving perceptual stealth (via blending, sample-specific generation, or frequency concealment) often increases fragility to aggressive physical transformation or heavy augmentation, so rigorous evaluations must report both invisibility and persistence under real-world image pipelines.

\subsubsection{Semantic Triggers}  
Semantic triggers rely on naturally occurring, semantically meaningful attributes (e.g., clothing colour, accessories, background objects, or contextual co-occurrences) to activate a backdoor, rather than on artificial stickers or pixel-level patterns. Because poisoned samples can retain plausible labels, semantic-trigger attacks are often label-consistent and highly stealthy: early clean-label and label-consistent strategies showed how adversarial or generative modifications can embed trigger–target associations without obvious label/image mismatch \cite{turner2019label,saha2020hidden}. 

More recent work develops explicit semantic-trigger pipelines that exploit attributes or context as activation cues. For example, Wang et al. \cite{wang2023versatile} propose VSSC triggers that automatically select visible, semantic, sample-specific, and physically compatible objects as robust in-the-wild triggers, and Chen et al. \cite{chen2017targeted} illustrated how everyday accessories (e.g., eyeglass frames) can be used as physical semantic triggers to subvert face-recognition systems. Barni et al. \cite{barni2019new} and Souri et al. \cite{souri2022sleeper} demonstrate related hidden-attribute or label-consistent strategies that generalize semantic triggers to settings where labels are not corrupted or where triggers are concealed within plausible image variants. 

On dense-prediction tasks, Li et al. \cite{li2021hidden} propose a fine-grained hidden backdoor for semantic segmentation that relabels only object-level pixels to the attacker-specified target while leaving other pixels intact, enabling precise, low-rate manipulation of segmentation outputs without degrading global performance. Because semantic triggers align with natural data patterns, simple input filtering or patch removal is often ineffective. Effective detection and mitigation instead require richer representation or causality-aware analyses, such as SODA \cite{sun2024neural}, along with provenance-based measures that can separate spurious attribute-label correlations from genuine semantics.

\subsubsection{Input-transformation Triggers}  
Input-transformation triggers activate backdoors by applying global or non-local alterations to training examples rather than by stamping a localized sticker; the poisoned samples therefore remain visually or statistically close to benign data while nevertheless inducing a persistent trigger to target mapping. Early work showed that low-opacity blending and mixed-image overlays can embed reliable triggers with reduced conspicuity \cite{chen2017targeted}, and follow-up studies explored sample-specific invisible noises produced by encoder–decoder or steganographic generators to defeat defenses that assume a single, universal pattern \cite{li2020invisible}. 

Other methods constrain the perturbation norm or directly optimize universal perturbations to minimize perceptual footprint while retaining attack potency \cite{turner2019label,DBLP:journals/www/HouHYKT23}. Beyond pixel-space manipulations, a number of works exploit natural image phenomena or preprocessing operations as triggers: Refool \cite{liu2020reflection} plants reflections as physically plausible triggers, WaNet \cite{nguyen2021wanet} uses imperceptible geometric warps to warp representations rather than add conspicuous pixels, and image-scaling attacks hide triggers that only appear after common resizing operations \cite{quiring2020backdooring}. 

A complementary line of research reframes triggers in the frequency domain: careful embedding of energy into selected spectral bands produces diffuse, low-amplitude pixel perturbations that are hard to detect yet coherent in feature space \cite{zeng2021rethinking,wang2021backdoor}. Hidden or clean-label input-transformation attacks (including gradient-matching and data-selection strategies exemplified by Sleeper Agent \cite{souri2022sleeper}) further raise stealth by avoiding label corruption and by crafting poisons that resemble in-distribution data while still steering learned features. 

Because many of these transformations exploit global statistics, preprocessing pipelines or learned encoders, they challenge defenses that rely on detecting localized anomalies; robust mitigation therefore requires a mixture of preprocessing-hardened training, representation-space auditing, and threat-aware evaluation under realistic augmentations.

\subsection{Labeling Strategy}
\label{sec:labeling-strategy}

The labeling strategy of a backdoor attack describes whether and how poisoned examples are assigned labels during training, and it strongly influences both detectability and the range of viable defenses. Two broad families dominate the literature: \emph{dirty-label} attacks, where poisoned inputs are paired with attacker-chosen target labels, and \emph{clean-label} attacks, where poisoned inputs retain semantically correct labels and the trigger is engineered to alter feature representations rather than label text. Each family presents distinct trade-offs between ease of crafting, stealth, and robustness under typical data-curation pipelines.

\subsubsection{Dirty-label Attacks}  
Dirty-label attacks are the canonical and most straightforward labeling strategy for backdoor injection: the adversary inserts trigger-bearing examples into a training corpus and deliberately assigns them the attacker-chosen target label, creating an explicit trigger to label mapping that the model learns during optimization. The classical demonstration is BadNets \cite{gu2019badnets}, which showed that stamping a small, consistently-placed patch onto a small fraction of training images and relabeling them suffices to induce highly reliable misclassification on patched inputs while preserving clean accuracy. 

Subsequent work generalized this recipe along multiple axes: Chen et al. \cite{chen2017targeted} formalized targeted data-poisoning methods and illustrated physical instantiations (e.g., eyeglass-frame-like triggers for face recognition), and Li et al. \cite{li2021hidden} extended the dirty-label concept to dense prediction by relabeling only object pixels to mount fine-grained segmentation backdoors. Other works in semantic segmentation domain that utilize dirty-label attack strategy include \cite{lan2024influencer,mao2023object}. Patch-style dirty-label attacks have also been adapted to object detection and other deployed systems, where localized stickers or printed patterns can cause targeted misdetections or cloaking in real-world settings \cite{chan2022baddet,ma2022dangerous}. 

While dataset-poisoning at scale is conceptually simple, practical pipelines such as federated learning and large-scale data curation amplify the threat: a malicious client or poisoned data source can introduce dirty-label examples that evade manual inspection, and model or client-level aggregation may further obscure their presence \cite{bagdasaryan2020backdoor}. Importantly, dirty-label poisons are often the easiest family to detect using dataset-sanitization or representation-based filters (e.g., spectral signatures \cite{tran2018spectral}, activation clustering \cite{chen2018detecting}), but in large, noisy, or weakly curated datasets many poisoned instances may escape such defenses and still permit highly effective attacks.

\subsubsection{Clean-label and Feature-collision Attacks}  
Clean-label attacks aim for stealth by ensuring poisoned examples remain correctly labeled to human annotators, forcing the model to learn the backdoor through subtle representation interference rather than via an explicit trigger to label mapping. Early, seminal work formalised the *feature-collision* idea: crafting poisons so that their deep features collide with those of a chosen target causes the target to be mapped to the poison class after training, while the poison images themselves remain visually consistent with their (clean) labels \cite{shafahi2018poison}. 

Subsequent advances improved transferability and practicality: polytope-style and ensemble-based poisoning produces more transferable poisons that entrain targets inside convex regions of feature space \cite{zhu2019transferable}, and MetaPoison \cite{huang2020metapoison} applied meta-learning to approximate the bi-level optimization needed for strong, general-purpose clean-label poisons that work even when models are trained from scratch. Other work developed *hidden-trigger* or gradient-matching approaches (e.g., Sleeper Agent \cite{souri2022sleeper}) that scale to large datasets and black-box settings by matching training dynamics rather than explicit feature collisions. 

Clean-label strategies have also been adapted to different modalities and threat constraints: clean-label video backdoors use universal temporal triggers to attack action recognition models \cite{zhao2020clean}, while Narcissus \cite{zeng2023narcissus} demonstrated practical clean-label attacks that require only limited knowledge of representative target examples and achieve high success at very small poisoning rates. A similar but more advanced method where authors utilize alternating surrogate-trigger optimization strategy is introduced in \cite{huynh2024combat}. 

Finally, multi-target clean-label attacks like FFCBA introduce class-conditional feature-based perturbations through autoencoder frameworks, allowing simultaneous targeting of multiple labels while retaining natural feature coherence in the poison samples \cite{yin2025ffcba}. These methods show that clean-label and feature-collision attacks can be highly stealthy and effective, but they typically demand more elaborate optimization, stronger white-box assumptions, or surrogate data; defenses therefore need to target representation-level anomalies (feature distributions or training dynamics) rather than simple label-sanity checks.

\subsection{Representation Stage}
\label{sec:representation-stage}

The \emph{representation stage} classification addresses \emph{where} in a model's learned latent space the backdoor is encoded and \emph{how} the trigger manifests in internal activations. Whereas earlier dimensions (injection stage, trigger design, labeling strategy) emphasize how poisoned inputs are produced, the representation-stage view highlights the adversary’s objective in shaping feature distributions or neuron activations so that the presence of a trigger produces a predictable trajectory through the network’s latent space. This perspective is particularly useful when comparing attacks that are indistinguishable at the pixel level but differ fundamentally in the depth, breadth, or semantic scope of their influence on learned representations. In what follows we divide representation-stage attacks into five subcategories and summarize the standard works and trends for each.

\subsubsection{Instance-specific Representation Attacks}  
Instance-specific representation attacks tie the backdoor to a narrowly defined region of feature space or to per-sample signals so that only a small, carefully-crafted set of inputs will activate the malicious behaviour. These attacks break a central assumption of many classical defenses, that a single, universal trigger or a small set of repeated triggers underlies the trojan and they are therefore substantially harder to detect using spectral \cite{tran2018spectral}, clustering \cite{chen2018detecting}, or single-trigger reverse-engineering techniques \cite{wang2019neural}. Representative examples include the Input-Aware Dynamic Backdoor Attack \cite{nguyen2020inputaware}, which trains an input-conditioned generator to produce unique triggers per sample and enforces non-reusability via cross-trigger constraints; the resulting trojaned models achieve high attack success while evading Neural-Cleanse \cite{wang2019neural}, STRIP \cite{gao2019strip} and other fixed-trigger detectors. 

Invisible, sample-specific attacks adopt a steganographic approach: Li et al. \cite{li2021invisible} use an encoder–decoder to embed short, attacker-chosen bit-strings as imperceptible, per-image perturbations. These encoded bit-strings, whicha are not human-readable text but numerical payloads, serve as triggers that are visually benign yet learned by the network during training to activate the attacker’s chosen label. Instance-specific strategies also appear in the pre-trained-encoder threat model: BadEncoder \cite{jia2022badencoder} and related works show that an attacker who compromises or crafts a pre-trained encoder can cause downstream classifiers to inherit encoder-level backdoors that are effectively instance or feature-specific across multiple tasks \cite{shen2021backdoor}. 

Because these attacks are crafted to minimize overlap with the clean data distribution and to avoid repeated or outlier-like patterns, defenses cannot rely solely on single-layer or single-trigger scans. Promising approaches include representation-space auditing and active separation, such as ASSET \cite{pan2023asset}, which encourages distinguishable behavior between clean and poisoned samples; encoder-scanning techniques, such as DECREE \cite{feng2023detecting}, which directly probe pre-trained encoders; and hybrid methods that combine active data perturbation with multi-layer signal aggregation \cite{hayase2021spectre}. Nevertheless, instance-specific backdoors remain among the most challenging threats, as they are sample-specific, covert, and transferable across downstream tasks. Effectively defending against them requires approaches that are both task-aware and capable of probing model behavior across multiple representation levels.

\subsubsection{Class-wide / Manifold-level Attacks}
Class-wide representation attacks operate at the manifold level: they aim to remap poisoned or triggered inputs so that their latent embeddings collapse into the target-class distribution, making them indistinguishable from genuine target examples. This mechanism underlies canonical backdoors such as BadNets \cite{gu2019badnets} and blended-patch poisons \cite{chen2017targeted}, where triggered inputs are absorbed into the target-class manifold in deep feature space. Because of this collapse, clustering or distance-based anomaly defenses struggle to separate poisoned and clean samples. 

In the transfer-learning setting, poisoned encoders can carry these manifold-level associations into downstream tasks even after fine-tuning, broadening their impact to diverse modalities and datasets \cite{shen2021backdoor,wei2023aliasing}. Recent studies further show that transformer architectures are particularly sensitive to such representation-level poisoning, as attention-driven encoders can entangle spurious trigger–label correlations into the learned embedding space \cite{gong2024megatron}. Compared to instance-specific triggers, class-wide attacks prioritize universality and persistence, often achieving higher generalization at the cost of easier cross-sample detectability when sufficiently strong representation-level audits are applied.

\subsubsection{Neuron / Parameter-hijacking Attacks}  
Neuron and parameter-hijacking attacks bypass the input space entirely, embedding backdoors by directly manipulating a model’s internal neurons, parameters, or micro-architecture. These modifications cause malicious behavior to be triggered by small, often subtle, backdoor triggers. Early and influential demonstrations showed that a malicious builder can ``trojan'' a network by surgically modifying weights or inserting trigger-sensitive sub-networks into a released checkpoint, producing a model that behaves normally on benign data but fires on attacker-chosen cues \cite{liu2017trojaning}. Complementary work formalised targeted weight perturbations, demonstrating that carefully crafted, low-magnitude changes to specific layers or weight blocks can implant persistent misclassification behaviors with minimal impact on clean accuracy \cite{dumford2020backdooring}. At the neuron level, ABS (Artificial Brain Stimulation) \cite{liu2019abs} explored how to locate and characterize units that respond selectively to trojan triggers by actively probing internal neurons, underscoring that backdoors often manifest as localized neuron sensitivities that can be exploited or inspected.  

Parameter-space hijacks take multiple operational forms. Supply-chain compromises and poisoned checkpoints (e.g., BadEncoder-style attacks \cite{jia2022badencoder}) show that encoding a backdoor in a pre-trained encoder causes downstream classifiers to inherit the trojan after fine-tuning, effectively weaponizing the model-provenance channel \cite{shen2021backdoor}. Layerwise weight-poisoning further demonstrates that attackers can target particular layers so that standard fine-tuning or pruning does not readily remove the trojan \cite{li2021backdoor}. Hardware and memory-level faults (e.g., targeted bit-flip \cite{dong2023one} or Rowhammer-style attacks \cite{cai2024wbp}) form another pragmatic pathway: flipping only a handful of bits in model storage can convert a benign checkpoint into a trojaned one, creating a stealthy, persistent compromise that is difficult to trace. Because these attacks act in parameter-space, they evade defenses that focus strictly on training-data anomalies and demand complementary integrity controls cryptographic provenance, parameter-space anomaly detection, neuron-stimulation audits, and targeted repair (fine-pruning, constrained re-training) to detect and remediate hijacked models \cite{guo2019tabor,Liu2018FinePruningDA}. Parameter-hijacking, thus represents a high-impact threat class: it requires fewer assumptions about access to the training pipeline, is naturally suited to supply-chain and federated contexts, and can be designed to persist through many downstream transformations.

\subsubsection{Distributed / Multi-layer Encoding}  
Distributed or multi-layer encoding attacks spread the backdoor signature across several representation levels or into global statistics (e.g., style, frequency bands, or geometric fields) rather than localizing it to a small pixel region or individual neurons. By dispersing the trigger’s footprint, these methods intentionally defeat detection strategies that inspect single layers, localized activations, or simple spectral signatures. Representative examples include warping-based attacks such as WaNet \cite{nguyen2021wanet}, which use imperceptible geometric deformations that alter mid-level feature maps across many layers and remain largely transparent to human inspection and simple layer-wise probes; frequency-domain attacks \cite{wang2021backdoor,zeng2021rethinking} that inject trigger energy into selected spectral bands so the pixel-space perturbation is diffuse yet coherent in feature space; and style or feature-space encodings that treat style (texture, makeup, global coloration) as the trigger modality. 

Notable style/feature-space works include the study on style-based feature perturbations for adversarial attack \cite{xu2021towards} and recent face-recognition backdoor attacks (e.g., MakeupAttack \cite{sun2024makeupattack}) that use realistic makeup-style transfer to implant backdoors in a way that is semantically natural and robust to many common augmentations. Because the malicious signal is embedded across layers and tied to distributed statistics, reliable detection requires multi-level or statistic-aware auditing (for example, multi-layer representation scans, frequency-aware checks, and robust covariance estimators such as SPECTRE \cite{hayase2021spectre}) rather than single-point inspections. These attacks therefore motivate defenses that aggregate signals across layers and domains, combine frequency/feature-space checks, and verify model provenance and parameter integrity to increase the chance of exposing distributed encodings.

\subsection{Target Tasks}
\label{sec:target-tasks}
Backdoor attacks in computer vision manifest differently depending on the \emph{target task}, as each task involves distinct model architectures, data modalities, and performance objectives. While image classification has traditionally been the primary focus due to its foundational nature and relative simplicity, recent work increasingly addresses more complex and safety-critical tasks such as object detection, semantic segmentation, and face recognition. The vulnerability of these tasks to backdoor attacks varies according to how triggers affect spatial, temporal, or semantic features within the data. This subsection categorizes backdoor attacks based on their intended target task, offering insights into the specific challenges and attack strategies relevant to each domain.

\subsubsection{Image Classification}  
Image classification is the most extensively studied target task in backdoor research owing to its central role in computer vision and the availability of standard benchmarks (e.g., CIFAR, ImageNet) that make evaluation and reproduction straightforward. The research lineage begins with BadNets \cite{gu2019badnets}, which showed that stamping a small visible patch onto a modest fraction of training images and relabelling them yields very high attack success rates while preserving clean accuracy. 

Subsequent work diversified and hardened the classification threat model: blending and low-opacity overlays reduce perceptual conspicuity \cite{chen2017targeted}; label-consistent and feature-collision techniques (PoisonFrogs \cite{shafahi2018poison}, MetaPoison \cite{huang2020metapoison}, Polytope-style poisons \cite{zhu2019transferable}) craft visually plausible poisons that evade simple label-sanity checks; invisible and sample-specific triggers (input-aware \cite{nguyen2020inputaware}, steganographic \cite{li2021invisible}) produce per-instance perturbations that break fixed-trigger detectors; and geometric/transform/frequency-based triggers (WaNet \cite{nguyen2021wanet}, frequency-perspective attacks \cite{wang2021backdoor,zeng2021rethinking}) move the malicious signal into non-obvious domains of the processing pipeline. 

The threat has further broadened with attacks on pre-trained encoders and transfer learning: backdoored encoders and checkpoints can carry manifold-level associations into downstream classifiers after fine-tuning \cite{jia2022badencoder,shen2021backdoor}. Because many defenses (e.g., spectral signatures \cite{tran2018spectral}, Neural Cleanse \cite{wang2019neural}, STRIP \cite{gao2019strip}) were developed and evaluated primarily on classification benchmarks, advances in classification attacks both drive and reveal gaps in defensive techniques.

\subsubsection{Object Detection}
Backdoor attacks pose a significant security threat to object detection models, which are critical for safety-sensitive applications like autonomous driving, robot vision, and video surveillance \cite{ma2022dangerous,cheng2023attacking,doan2024credibility,li2023badlidet}. These attacks involve maliciously poisoning training data such that the infected model performs normally on benign inputs but exhibits attacker-specified misbehavior when a hidden trigger is present. The malicious behaviors can include `cloaking objects' (making them disappear from detection) \cite{ma2022dangerous,chan2022baddet,zhang2022towards}, `misclassifying objects' (e.g., a car as a person) \cite{zhang2022towards,yin2024physical}, or causing `object disappearance' \cite{qian2023robustbackdoorattacksobject,10647450,li2023badlidet}, among other effects like object generation \cite{cheng2023attacking,10647450}. 

Attackers can use various triggers, ranging from digital patterns like chessboard triggers or white patches, to physical and natural objects such as T-shirts, Post-it Notes, flower stickers, RGB stickers, Target stickers, cargo carrier bags, or exercise balls. Some advanced attacks even employ `invisible triggers' \cite{li2024twintriggergenerativenetworks} or manipulate detection based on `temperature modulation' \cite{yin2024physical}, making them difficult to detect by human inspection. These attacks have been validated across various scenarios, including data outsourcing, model outsourcing, and pre-trained model fine-tuning, and even in the context of Self-Supervised Learning (SSL). 

\subsubsection{Semantic Segmentation}
Backdoor attacks in semantic segmentation present unique challenges due to pixel-wise classification and contextual dependencies \cite{li2021hidden,mao2023object,lan2024influencer}. Early methods like the Fine-Grained Backdoor Attack (FGBA) \cite{li2021hidden} targeted specific object pixels but suffered from low attack success rates and contextual inconsistencies. The Object-Free Backdoor Attack (OFBA) \cite{mao2023object} flexibility by allowing dynamic selection of attacked classes and trigger positions during inference, requiring digital triggers placed on victim pixels. The Influencer Backdoor Attack (IBA) \cite{lan2024influencer} introduced small, natural triggers on non-victim pixels to indirectly cause misclassification, using strategies like Nearest Neighbor Injection and Pixel Random Labeling to leverage broader context. Despite practical advances, IBA’s triggers remain conspicuous and detection-prone. These developments underscore the growing security concerns of backdoor threats in segmentation and highlight the need for robust, context-aware defense mechanisms.

\section{Backdoor Defenses}
\label{sec:backdoor-defenses}

Defending against backdoors requires a similarly multi-faceted taxonomy to the one we used for attacks: effective countermeasures depend on \emph{where} in the machine-learning pipeline they are applied, \emph{what} they are intended to achieve (detection, mitigation, repair or provable robustness), the \emph{trust and access} assumptions they require (white-box vs. black-box, access to training data or only to a deployed model), and the \emph{target task} and threat model they address (classification, detection, segmentation, pre-trained encoders, federated learning, etc.). Organising defenses along the same high-level axes as the attacks (injection stage, trigger type, labeling strategy, representation stage, and target task) makes it straightforward to (i) map each defensive technique to the specific families of attacks it can plausibly counter, and (ii) expose gaps where particular attack classes remain largely undefended.  

In this section we follow that mirrored taxonomy. For each pipeline stage we first summarize the defense objectives and threat assumptions common to that stage, then review representative methods and the principal evaluation criteria used in the literature. Concretely, Section~\ref{sec:def-inject} surveys \emph{pre-training and training-time} interventions (dataset sanitization, robust training and data-level detection); Section~\ref{sec:defenses-by-trigger} covers \emph{trigger-aware} techniques (patch/localiser detectors, frequency or transformation-aware filters, and sample-specific/randomized defenses); Section~\ref{sec:def-labeling} discusses \emph{labeling-strategy} mitigations (dirty-label sanitizers, clean-label/feature-collision detectors, and adaptive labeling schemes); Section~\ref{sec:defenses-representation} treats \emph{representation and parameter} defenses (activation auditing, encoder scanning, multi-layer/statistic-based detection, and parameter-integrity checks); and Section~\ref{sec:defenses-target-tasks} examines task-specific defenses and evaluation practices for classification and object detection.  


\subsection{Injection Stage}
\label{sec:def-inject}
Defenses targeted at the \emph{injection stage} operate before or during training and aim to prevent poisoned samples from influencing the learned model. These pre-training countermeasures typically try to (a) detect and remove suspicious examples from the training corpus, (b) purify or repair suspicious samples so they no longer carry trigger features, or (c) harden the training procedure so that any remaining poisons have reduced influence. Because they intervene early in the pipeline, injection-stage methods can provide strong, proactive protection for downstream models and supply-chain consumers; their effectiveness, however, depends strongly on the defender’s access to training data, trusted validation sets, intermediate activations, and the threat model (e.g., clean-label vs dirty-label, sample-specific triggers). 

\subsubsection{Data Sanitization (pre-training)}
A dominant pre-training defense strategy is \emph{dataset sanitization}: inspect training examples (often via learned intermediate representations) and filter or repair those that appear anomalous. Early representation-based filters exploited the observation that poisoned examples often leave a detectable footprint in deep features. Tran et al. \cite{tran2018spectral} proposed a PCA/spectral approach that identifies a top principal direction (a ``spectral signature'') along which poisoned examples separate from clean ones; removing the outlying samples substantially reduces backdoor effectiveness . Building on this idea, Activation Clustering \cite{chen2018detecting} clusters per-class hidden activations and flags small, tight clusters as candidate poisons, allowing defenders to remove or relabel suspicious subsets without requiring a trusted clean set. Subsequent work improved robustness and applicability: SPECTRE \cite{hayase2021spectre} detects poisoned data by first performing robust estimation of the clean distribution’s mean and covariance, then applying a whitening transform that decorrelates features and equalizes variance across directions. This normalization amplifies the subtle spectral signatures introduced by backdoor samples, making them separable in cases where standard PCA fails. 

Complementary approaches avoid per-sample filtering and instead ask the model whether it is poisoned: Universal Litmus Patterns (ULPs) \cite{kolouri2020universal} feed diagnostic patterns to the network and use the response distribution to flag corrupted datasets quickly and with few forward passes. For clean-label or feature-collision attacks, defenses that operate in representation space such as Deep k-NN use nearest-neighbour statistics to detect anomalous feature alignments that are otherwise label-consistent \cite{peri2020deep}. More recently, purification approaches like DataElixir \cite{zhou2024dataelixir} apply generative or diffusion-based image restoration to strip potential trigger features and recover benign content, thereby converting poisoned samples back to useful data instead of discarding them. In parallel, robust optimization and data-selection methods (e.g., SEVER \cite{diakonikolas2019sever}, iterative trimmed-loss \cite{shen2019learning}) aim to make training itself resistant to a small fraction of outliers by iteratively identifying and down-weighting anomalous losses. 

\subsubsection{Training-time Interventions}  
Training-time defenses aim to prevent backdoors from being learned in the first place by changing the optimization dynamics, the loss landscape, or the dataset usage during training. A central inspiration for these methods is the empirical observation that poisoned examples behave differently from clean samples during training (e.g., they often produce low losses early or exhibit distinct robustness properties), and that disrupting these dynamics can isolate or unlearn backdoor correlations. Anti-Backdoor Learning (ABL) \cite{li2021anti} formalizes this idea via a two-stage training scheme that (1) uses gradient-ascent style steps early to isolate low-loss (likely poisoned) samples and (2) breaks the trigger to target association by unlearning those samples in later epochs; ABL demonstrates that a carefully designed training schedule can yield models trained on poisoned dataset that behave like models trained on clean data. 

Follow-ups and variants refine the isolation step or integrate meta-learning: Adaptively Splitting Dataset (ASD) \cite{gao2023backdoor} dynamically partitions the training set into candidate clean and polluted pools using loss-guided and meta-learning updates, improving robustness during the main training phase, while Progressive Isolation (PIPD) \cite{chen2024progressive} iteratively refines isolation to reduce false positives when identifying poisons. Another complementary family leverages robust optimization primitives: differential privacy mechanisms (e.g., DP-SGD and PATE) limit any single sample’s influence during updates and have been shown, when properly tuned to reduce backdoor success, though effectiveness depends strongly on hyperparameters and can trade off clean accuracy \cite{10.1007/978-3-031-65172-4_20}. 

Adversarially-inspired training methods connect adversarial robustness and backdoor mitigation: several works examine adversarial training or adversarial fine-tuning as a countermeasure, reporting that spatially-aware adversarial training and hybrid strategies can suppress certain patch and transformation triggers, and that adversarial fine-tuning (training on adversarial examples of an infected model) can help erase triggers from a compromised checkpoint \cite{gao2023effectiveness,Mu2022AdversarialFF}. Building on this idea, Adversarially Robust ABL (A-ABL) \cite{zhao2024adversarially} combines dataset splitting with adversarial training to produce models that are both backdoor-free and adversarially robust. 

In practice, training-time interventions are attractive because they require no post-hoc inspection of checkpoints, but they typically require control of the training pipeline, careful hyperparameter tuning, and assumptions about poisoning rate or loss-gap behaviors; sophisticated or adaptive attackers (sample-specific triggers, extremely low poisoning rates, or triggers designed to mimic clean-sample dynamics) can still evade these methods, so they are best used in concert with complementary sanitization and model-level checks.

\subsubsection{Post-training Model Inspection and Repair}  
Post-training defenses inspect a delivered model to determine whether it contains a backdoor and when possible, repair the checkpoint so it no longer responds to triggers. A widely used strategy is trigger reverse-engineering: methods such as Neural Cleanse  \cite{wang2019neural} search for a minimal input perturbation that forces the network to predict a given label; an anomalously small reverse-engineered trigger is taken as evidence of a backdoor and can be used to neutralize the attack (e.g., by patching or pruning). 

Complementary approaches probe internal activations: Activation Clustering \cite{chen2018detecting} groups hidden-layer activations to separate poisoned from clean samples and identify candidate poisons without requiring access to the training set, while ABS (Artificial Brain Stimulation) \cite{liu2019abs} actively stimulates neurons to discover trojan-related activation patterns and reconstruct triggers from neuron responses. Once a trigger or suspicious neurons are identified, a common repair pipeline uses targeted pruning and selective fine-tuning (e.g., Fine-Pruning \cite{liu2018fine}) or model patching techniques that directly adjust model parameters to erase the malicious behavior while preserving clean accuracy \cite{liu2024mudjacking}. 

Black-box variants such as DeepInspect \cite{chen2019deepinspect} and TABOR \cite{guo2019tabor} combine trigger synthesis with anomaly scoring and model patching to work with limited model access. These post-training techniques are practical for model consumers who cannot retrain from scratch, but they face limitations: reverse-engineering can fail for large, sample-specific, or distributed triggers; pruning-aware or adaptively trained backdoors can evade neuron-based scans; and optimization-heavy inspection (e.g., per-label trigger search) can be computationally expensive, motivating hybrid pipelines that combine fast detectors with selective, deeper inspection for flagged checkpoints.

\subsubsection{Test-time Input Filtering and Mitigation}  
Runtime defenses operate at inference time and aim to detect or neutralize trigger-bearing inputs before they can activate a backdoor, making them especially attractive when model retraining or checkpoint inspection is infeasible. A common and influential strategy is perturbation-based detection: STRIP \cite{gao2019strip} intentionally mixes an incoming image with a variety of perturbations and measures the entropy of the model’s predictions, low entropy indicates a likely trigger sample because the backdoor forces a stable, input-independent output. Complementing entropy tests, input-purification methods such as Februus \cite{doan2020februus} surgically remove suspicious regions (via trigger extraction and inpainting) to restore benign content and thus neutralize many patch-style trojans without requiring model updates. 

Saliency and interpretability-driven detectors (e.g., SentiNet \cite{Chou2020SentiNetDL}) leverage model explanation tools to localize highly salient contiguous regions and then test their influence across inputs, enabling agnostic detection of localized, reusable triggers in the physical world. More recent work (TeCo \cite{liu2023detecting}) exploits differences in corruption-robustness: backdoor-triggered inputs tend to show inconsistent prediction transitions under common image corruptions compared to clean inputs, allowing test-time detection using only hard-label outputs and no extra clean data. Runtime defenses are attractive for their practicality and modest deployment requirements, but they incur inference latency, can produce false positives on legitimate unusual inputs, and may be vulnerable to adaptive attackers who design triggers to break the perturbation or purification heuristics; thus, they are most effective when combined with upstream (sanitization/training) or downstream (model inspection) measures as part of a defense-in-depth strategy. 

\subsection{Trigger type}
\label{sec:defenses-by-trigger}
Organizing defenses by assumed trigger type is practically useful because many countermeasures rely on structural assumptions about the trigger (locality, visual saliency, spectral footprint, semantic plausibility, or per-sample randomness). A detector or repair method tuned for compact, visible patches will perform well against sticker-style backdoors but can fail catastrophically for blended, frequency-domain, or sample-specific triggers; conversely, representation-level or transformation-aware defenses are more general but often costlier and less precise. In what follows we review defenses that explicitly assume a compact, localized trigger, a common case in many physical-world attacks and examine their techniques, strengths and failure modes.

\subsubsection{Patch Detectors}  
Patch and sticker-focused defenses target triggers that are spatially compact and (typically) reusable across inputs. A prominent family of methods uses model explanations and saliency to localize suspicious contiguous regions: SentiNet \cite{Chou2020SentiNetDL} leverages saliency maps to find highly salient regions and then tests whether transplanting those regions onto other inputs forces misclassification, thereby both localizing and confirming physical, reusable patches. Reverse-engineering approaches seek the smallest perturbation (mask or pattern) that, when applied to a set of inputs, forces them into a particular label; Neural Cleanse \cite{wang2019neural} operationalizes this idea by optimizing per-label minimal triggers and flagging labels with abnormally small reverse-engineered patterns as likely backdoored, enabling downstream mitigation via masking/pruning. 

Complementary repair-oriented techniques combine localization with input restoration: Februus \cite{doan2020februus} first extracts a suspected trigger region using attention/heatmap signals and then inpaints the region to neutralize the trigger before classification, effectively removing many patch-style trojans at inference time. Black-box or limited-access pipelines adapt these ideas: DeepInspect \cite{chen2019deepinspect} and similar frameworks synthesize candidate triggers from model outputs to score and then mitigate suspicious labels without full white-box access \cite{guo2019tabor}. Neuron-stimulation methods such as ABS \cite{liu2019abs} probe internal units to identify neurons that respond selectively to triggers and then use those neurons to reconstruct or localize patch triggers for repair. 

While these approaches can be highly effective against visible, fixed-location or reusable patches, they share common limitations: (1) they assume trigger locality and reusability, so they are often bypassed by blended, distributed, transformation-based, or sample-specific triggers; (2) reverse-engineering and inpainting can be computationally expensive and suffer false positives on legitimate but unusual inputs; and (3) adaptive attackers can design triggers that minimize saliency or produce large reverse-engineered masks, reducing anomaly signals. In practice, patch/localizer defenses work best as part of a layered pipeline (quick localization + lightweight mitigation + deeper inspection for flagged models) rather than as standalone guarantees. 

\subsubsection{Blend or Invisible-perturbation Defenses}  
Detecting and mitigating blended or invisible triggers; those spread across pixels or embedded in frequency bands, poses a substantially harder problem than localized patch detection because such triggers leave only subtle footprints in pixel space and often mimic in-distribution variations. Defense approaches therefore trend away from purely spatial heuristics and toward (1) spectral and frequency analysis, (2) representation-level anomaly detectors, (3) input purification/denoising, (4) model-level litmus and consistency tests, and (5) robust training that limits any single sample’s influence. Spectral-signature methods and follow-ups (e.g., Tran et al.’s spectral signatures \cite{tran2018spectral}, and Zeng et al.’s frequency-perspective analysis \cite{zeng2021rethinking}) inspect principal components or selected frequency bands to expose concentrated trigger energy that is invisible in pixel space but observable in transform domains \cite{wang2021backdoor}. 

Representation-level detectors such as Activation Clustering \cite{chen2018detecting} and SPECTRE \cite{hayase2021spectre} improve resilience by using robust covariance estimation and clustering in deep feature space to flag subtle, low-amplitude poisons that evade pixel-level filters. Input purification and denoising approaches attempt to remove potential trigger perturbations before inference, recent work uses generative or diffusion-based purification to reconstruct benign content from suspected poisoned samples, which can neutralize many invisible triggers without needing to identify them explicitly \cite{zhou2024dataelixir,doan2020februus}. Universal litmus tests and consistency checks offer fast dataset or model-level alarms by probing model responses to designed probes or corruptions rather than searching for per-sample anomalies \cite{kolouri2020universal,liu2023detecting}. 

Finally, training-time robustification (differential privacy, Anti-Backdoor Learning variants) reduces the model’s capacity to memorize tiny, sample-specific perturbations, making invisible triggers harder to implant or transfer \cite{10.1007/978-3-031-65172-4_20,li2021anti}. All these techniques improve coverage against blended triggers compared to patch-centric methods, but they have trade-offs: frequency and feature detectors can be bypassed by adaptive attackers who disperse energy differently or craft sample-specific triggers; purification may alter benign content and incur computational cost; and robust training often degrades clean accuracy or requires access to the training pipeline. As a result, blended-trigger defense research emphasizes layered strategies, combining spectral scans, representation auditing, targeted purification, and robust training alongside attacker-aware evaluations to avoid over-claiming robustness.

\subsubsection{Semantic Triggers Based Defenses}  
Semantic triggers exploit meaningful, in-distribution attributes (e.g., accessories, background objects, color shifts) that correlate with target labels, so defenses that assume local or low-level anomalies often fail. A promising defense family therefore aims to \emph{disentangle} or \emph{deconfound} semantic attributes from causal label information so that spurious trigger to label associations are not learned or can be broken post-hoc. Causality-inspired Backdoor Defense (CBD) \cite{zhang2023backdoor} formalizes this idea by treating the trigger as a confounder and learning deconfounded representations via a two-model scheme that separates confounding and causal components, substantially reducing ASR while preserving clean accuracy in many settings. 

Related ideas that reshape the training signal include decoupling pipelines (self-supervised backbone + frozen encoder + semi-supervised fine-tuning) which reduce clustering of poisoned samples in feature space and thus limit feature-collision style attacks \cite{huang2022backdoor}. For contrastive and multimodal pretraining, semantic counterfactual augmentation (e.g., CleanerCLIP \cite{xun2024cleanerclip}) and knowledge-guided constraints (Semantic Shield \cite{ishmam2024semantic}) intervene during fine-tuning to disrupt trigger–target correlations by generating semantically plausible counterfactuals or enforcing semantic consistency, showing strong robustness against semantic triggers in VLMs and CLIP-style encoders. 

Prototype-guided and geometry-aware post-hoc sanitizers extend these ideas by penalizing activation displacements toward trigger-induced prototypes and fine-tuning the model away from suspect geometric movements in activation space \cite{amula2025prototype}. More generally, causality-inspired disentanglement using diffusion or generative models (e.g., CausalDiff \cite{zhang2024causaldiff}) offers a flexible toolkit for removing or isolating semantic confounders from representations. These semantic disentanglers are attractive because they directly target the root cause of semantic backdoors, but they typically require stronger assumptions (access to training/fine-tuning, auxiliary semantic priors or prototypes, or compute for counterfactual synthesis) and can suffer reduced efficacy against highly adaptive attackers who hide triggers inside legitimate semantic variation; consequently, they are best deployed together with dataset sanitization and robust evaluation under attacker-aware settings. 

\subsubsection{Transformation-aware Defenses}  
Transformation-aware defenses for backdoors exploit the fact that many trigger mechanisms; whether geometric warps, scaling artefacts, or subtle input perturbations, change their influence under controlled image transforms or preprocessing. A family of test-time detectors therefore checks for prediction `inconsistency' or `instability' across transformed views: TeCo \cite{liu2023detecting} evaluates the \emph{corruption-robustness consistency} of an input by applying a small suite of common corruptions and measuring deviations in prediction transition behavior; trigger samples exhibit markedly different transition severities compared to clean inputs and can thus be flagged without requiring model internals or extra clean data. 

Complementary runtime techniques apply targeted input-space operations to neutralize triggers: STRIP \cite{gao2019strip} uses random input perturbations (input mixing) and measures the entropy of predictions to detect inputs whose label is invariant under strong perturbations, a signature of input-agnostic backdoors, while Februus \cite{doan2020februus} extracts and inpaints suspicious regions (using attention/heatmap signals) to remove patch-like triggers before classification. Transformation-aware preprocessing can also be applied upstream to prevent trigger hiding via preprocessing attacks: image-scaling attacks, which can conceal triggers in downscaling stages, have motivated defenses that harden or detect malicious scaling (secure/resilient scaling algorithms and scaling-detection tests) to prevent the insertion or concealment of backdoors during preprocessing \cite{quiring2020adversarial,quiring2020backdooring}. 

There is growing interest in certifiable, transform-based guarantees: randomized-smoothing variants have been investigated for certifying robustness against localized patch-style triggers, though early studies show limited practical coverage for complex backdoors and motivate further research into smoothing schemes tailored to backdoor threat models \cite{wang2020certifying,levine2020randomized}. In practice, transformation-aware approaches are attractive because they can be deployed at inference with modest assumptions, but they incur runtime cost, may cause false positives on legitimately unusual inputs, and can be weakened by adaptive attackers who explicitly optimize triggers to be robust under the applied transforms; accordingly, they are most effective when combined with dataset sanitization and model-level inspection as part of a layered defense strategy.

\subsubsection{Sample-specific Defenses}  
Sample-specific or input-aware backdoors; where the trigger varies per input or is produced by a learned generator, pose one of the hardest defense challenges because they break the core assumption used by many detectors (that a small set of reusable, label-agnostic triggers exist). Defenses aimed at this threat therefore follow a few complementary strategies. One line attempts to reverse-engineer the trigger generator itself: by learning a compact generator that maps clean inputs to poisoned ones, defenders can detect the presence of content-aware backdoors and (when successful) use the learned generator to synthesise suspect samples for removal or model repair \cite{zheng2022detecting}. 

A related strategy is purification by proxy: Progressive Backdoor Erasing (PBE) \cite{mu2023progressive} leverages the empirical connection between adversarial examples and triggered inputs to iteratively purify a model without needing a clean extra dataset, it generates adversarial examples, fine-tunes the model on those examples to break trigger to label correlations, then uses the partially-purified model to identify cleaner training samples and repeat. Another recent effective family, works in the activation/feature space rather than the input space: BadActs (activation-space purification) \cite{yi2024badacts} detects abnormal neuron activation statistics and projects suspicious activations back toward learned clean activation intervals, thereby neutralizing both visible and feature-space (including sample-specific) triggers while preserving clean accuracy. 

Structural and training-time remedies supplement purification: Trap-and-Replace  \cite{wang2022trap} baits backdoors into a small, easy-to-retrain classification head (by adding an auxiliary reconstruction task) and then replaces that head using a small clean holdout, making removal feasible even when the trigger is content-aware. Ensemble and robustness-by-diversity approaches (e.g., BDEL \cite{xing2024bdel}) train or aggregate multiple diverse classifiers so that sample-specific triggers that hijack one model are less likely to transfer across the ensemble, improving resistance without relying on a trigger reconstruction step. In practice these defenses make different trade-offs: generator reverse-engineering can detect highly adaptive attacks but is optimization-heavy and may fail when the generator is complex; PBE and activation-space purification can work without a large clean dataset but require extra computation and (for some variants) a small clean validation set to estimate clean activation distributions; Trap-and-Replace and ensemble methods impose training-time changes or heavier compute but often give robust repairability. Because sample-specific triggers are explicitly designed to evade assumptions used by classic detectors, the most robust deployments combine detection, activation-level purification, and structural/ensemble measures and evaluate against contemporary dynamic/sample-specific attacks (e.g., Input-Aware Dynamic, sample-specific steganographic attacks, WaNet) in attacker-aware benchmarks.

\subsection{Labeling Strategy}
\label{sec:def-labeling}
The \emph{labeling strategy} dimension classifies backdoor attacks (and the corresponding defenses) according to how adversarial labels are handled during injection. This distinction matters because whether poisoned samples are deliberately \emph{mislabeled} (dirty-label) or retain their original, seemingly correct labels (clean-label / label-consistent) strongly influences both attack stealth and the set of effective countermeasures. Dirty-label attacks typically produce conspicuous label–input inconsistencies that can be exposed by label-sanity checks, influence analysis, or representation clustering, whereas clean-label and feature-collision attacks aim to remain plausible to human annotators by manipulating latent representations rather than surface labels, making them far harder to detect. Consequently, defenses organized along the labeling dimension emphasize different tools: data-centric and label-auditing techniques for dirty-label threats, and representation-level auditing, feature disentanglement, or certified/robust training for clean-label threats. In the sections that follow, we review methods tailored to each case and discuss their assumptions, typical success conditions, and known evasions.

\subsubsection{Dirty-label Focused Methods}  
Dirty-label backdoors; where poisoned inputs are deliberately relabeled to the attacker’s target are naturally amenable to defenses that inspect and correct label–image consistency before or during training. A common and effective strategy is to treat the detection problem as one of identifying label outliers in representation space: spectral-signature \cite{tran2018spectral} and activation-clustering methods \cite{chen2018detecting} exploit the empirical observation that, for many dirty-label attacks, poisoned examples form a separable subpopulation in deep feature space and can be flagged by PCA-based projections or by clustering hidden activations per class. 

Robust refinements such as SPECTRE \cite{hayase2021spectre} improve detection under weaker signal regimes by using robust covariance estimation and whitening to amplify subtle spectral footprints, widening applicability to lower poison rates and some adaptive variants. Complementary techniques borrow tools from data-debugging and label-error detection: influence-function analysis and confident-learning style pipelines (which estimate likely label errors via confusion/score statistics) can prioritize suspicious training examples for human review or automated removal, offering a pragmatic human-in-the-loop remediation path for dirty labels \cite{pmlr-v70-koh17a,northcutt2021confident}. 

Nearest-neighbor and prototype checks in feature space (e.g., Deep k-NN \cite{peri2020deep}) further help identify samples whose features are inconsistent with their assigned class, which is especially useful when label flips produce obvious representation mismatches. These approaches form a practical toolbox for defending against dirty-label poisoning: automated spectral/cluster filters for fast triage, influence and confident-learning scorers to prioritize human inspection, and representation-based nearest-neighbor tests to confirm or localize suspicious instances. Their limitations are also clear, sophisticated clean-label or sample-specific poisons that deliberately mimic target-class features can evade label-consistency signals, so dirty-label focused defenses are most reliable when combined with robust training and downstream model inspection for layered protection.

\subsubsection{Clean-label Defenses}
Clean-label and feature-collision poisons are among the most troublesome threats because the poisoned samples retain plausible labels and are crafted to sit on or near the legitimate data manifold, which defeats simple label-sanity checks or vanilla spectral filters. Defenses therefore concentrate on (1) detecting anomalous `feature' alignments rather than label inconsistencies, (2) making learned features more robust so that tiny, targeted perturbations cannot hijack downstream classifiers, and (3) using purification or active separation to break the delicate poison to target association. A practical and influential detection approach is ``Deep k-NN'' \cite{peri2020deep}, which flags training examples whose deep embeddings are inconsistent with their claimed class by examining nearest neighbors in representation space; Peri et al. report very high detection rates on classical feature-collision and convex-polytope attacks. 

Robust training and loss-based approaches, such as Anti-Backdoor Learning (ABL) \cite{li2021anti}, SEVER \cite{diakonikolas2019sever}, and iterative trimmed-loss minimization (ITLM) \cite{shen2019learning}, complement detection by reducing the model’s tendency to memorize small populations of adversarially chosen examples. Adversarially-aware pretraining or purposely widening inter-class feature gaps has recently been proposed to harden foundation encoders against transfer-learning style clean-label poisons \cite{blum2021robust}. Representation disentanglement \cite{zhang2023backdoor} and decoupled training pipelines \cite{huang2022backdoor} aim to remove or neutralize spurious attribute–label correlations that clean-label attacks exploit. 

Purification techniques, including recent diffusion/generative purifiers such as DataElixir \cite{zhou2024dataelixir} and active separation procedures like ASSET \cite{pan2023asset}, which intentionally induce behavioral offsets between clean and poisoned samples to promote separation, provide pragmatic pipelines that detect and remove poisoned points even in self-supervised and transfer-learning settings where many classic detectors fail. It is important to recognize practical limits: state-of-the-art clean-label attacks such as MetaPoison \cite{huang2020metapoison} and Narcissus \cite{zeng2023narcissus} are designed to transfer across architectures and hide in feature space, so no single countermeasure is universally reliable. Consequently, current best practice relies on a layered strategy combining feature-space detection (k-NN/prototype checks), robust/pretraining defenses, active separation or purification for flagged points, and careful adversary-aware evaluation using modern clean-label attacks.

\subsubsection{Adaptive Labeling Defenses}  
Adaptive labeling defenses are designed for threat models where the attacker may vary or obfuscate labeling behaviour (e.g., randomised label flips, label-noise injection, or dynamic clean-label strategies that try to mimic honest annotation). These defenses therefore avoid assuming a fixed label corruption pattern and instead rely on `adaptive' methods that learn which examples to trust, reweight, or quarantine during training, or that detect trojaned behavior at the model level via meta-analytic probes. A practical class of approaches adapts noisy-label learning techniques to the backdoor setting: ``Co-teaching'' trains two networks that select small-loss samples for each other, thereby preferentially training on likely-clean data and reducing the influence of poisoned labels \cite{han2018co}; ``MentorNet'' \cite{jiang2018mentornet} and related curriculum-learning or sample-weighting schemes learn a data-driven curriculum or weighting function that down-weights suspicious examples during optimization. Meta-learning and reweighting approaches explicitly optimize per-sample weights using a small trusted validation set (or via a meta-objective), making them effective when only limited clean data is available; the learning-to-reweight family demonstrates that this meta-gradient strategy can robustly reduce the impact of adversarially corrupted labels in practice \cite{ren2018learning}.

More recent backdoor-specific work formulates adaptive splitting or meta-detection pipelines: Adaptively Splitting Dataset (ASD) \cite{gao2023backdoor} dynamically partitions training data into candidate clean and polluted pools using loss-guided and meta-inspired updates, then treats the two pools differently during training to mitigate backdoor learning. Meta-model and meta-backdoor systems (e.g., MNTD \cite{xu2021detecting}) learn a higher-level binary detector (a meta-classifier) over model behavior or sample features by training on many simulated poisoned scenarios, enabling detection that generalizes across diverse and unforeseen attack strategies while operating with limited access assumptions. These adaptive method is attractive because it does not hard-code a single labeling assumption and can leverage small trusted datasets or shadow-model simulations to generalize to new attacks. Its limitations include sensitivity to the availability and representativeness of any trusted validation data, potential computational overhead from meta-optimization or shadow-model generation, and vulnerability to sophisticated attackers who explicitly optimize to mimic the loss/feature distributions of clean samples; accordingly, adaptive labeling defenses are most effective when combined with complementary sanitization, representation-level auditing, and adversary-aware evaluation.

\subsection{Representation Stage}
\label{sec:defenses-representation}
Defenses that operate at the \emph{representation stage} inspect and reason about a model's internal feature space and neuron activations rather than just the raw input or final output. Because backdoors ultimately manifest as anomalous structures in latent representations (e.g., a cluster of poisoned embeddings, an over-responsive neuron, or a coherent principal direction), representation-stage methods seek to (a) detect these anomalies by analyzing activations or embedding distributions, and (b) repair the model by removing or neutralizing the representation components that carry the backdoor. These techniques are especially valuable when one has white-box access to a checkpoint or when the defender can run diagnostic inputs through the model; however, they typically assume access to intermediate activations and, in many cases some labeled or trusted validation data for calibration.

\subsubsection{Activation Inspection}  
A major class of representation-stage defenses inspects neuron activations and deep embeddings to reveal signatures of poisoning. Early spectral and clustering observations show that poisoned examples frequently induce a coherent direction or a separable subpopulation in deep feature space; Tran et al. \cite{tran2018spectral} used spectral signatures to identify and remove such outliers by projecting activations onto principal directions and filtering extreme samples. Activation Clustering \cite{chen2018detecting} extends this idea by clustering per-class hidden activations and treating small, dense clusters as candidate poisons, enabling localization of suspicious training points without needing a trusted clean set. ABS \cite{liu2019abs} (Artificial Brain Stimulation) pursues an active strategy: it stimulates internal neurons to discover units that respond selectively to triggers, then uses these neurons to reconstruct or localize potential triggers for further analysis or removal. 

Fine-pruning \cite{Liu2018FinePruningDA} and related neuron-pruning repairs exploit the fact that trigger-induced behavior often depends on a small set of neurons; by pruning neurons that are disproportionately active on suspected triggered inputs and then fine-tuning, these methods can often remove backdoors while preserving clean accuracy. Robust covariance and whitening methods such as SPECTRE \cite{hayase2021spectre} improve detection when spectral signals are weak by using robust mean/covariance estimation to amplify subtle activation anomalies. Nearest-neighbor and prototype checks in embedding space (e.g., Deep k-NN \cite{peri2020deep}) complement cluster and spectral-based detectors by flagging samples whose representation neighborhood is inconsistent with their label, which is effective against both dirty-label and some clean-label feature-collision attacks. Finally, reverse-engineering approaches like Neural Cleanse detect anomalous label-specific triggers by solving a per-label minimal-trigger optimization; while Neural Cleanse \cite{wang2019neural} primarily searches in input space, its anomaly signal often correlates with atypical internal activations and thus bridges input- and representation-based inspection.

These activation-inspection techniques are powerful when their core assumptions hold (coherent poisoned clusters, localized neuron responses, or detectable spectral footprints) and when white-box access is available. Their limitations are also important: sample-specific or highly distributed triggers can avoid producing strong cluster/spectral signatures; adaptive attackers can distribute trigger influence across many neurons or craft poisons to mimic clean activation statistics; and some methods require careful calibration or a small trusted set to set thresholds. Consequently, activation inspection is most effective as part of a layered defense, used to triage checkpoints for deeper analysis, to guide targeted pruning/repair, and to provide interpretable evidence for forensic review.

\subsubsection{Feature-space Auditing and Encoder Scanning}  
As backdoors increasingly target pre-trained encoders and exploit subtle feature-space manipulations, defenses that directly audit embeddings and encoder behaviour have become essential. Feature-space auditing inspects embedding distributions, prototype distances, and activation statistics to reveal anomalous clustering or prototype drift indicative of poisoning; classic examples include spectral-signature and activation-clustering techniques that flag separable poisoned subpopulations in deep features \cite{tran2018spectral,chen2018detecting}, and nearest-neighbor / prototype checks (Deep k-NN \cite{peri2020deep}) that detect samples whose embeddings are inconsistent with their assigned label. 

Building on these ideas, more recent work explicitly targets pre-trained encoders and transfer/self-supervised settings: ASSET \cite{pan2023asset} actively induces behavioral offsets between clean and poisoned points to promote separation and robustly detects poisons across supervised, self-supervised and transfer-learning paradigms, while DECREE \cite{{feng2023detecting}} directly scans pre-trained encoders by searching for minimal trigger patterns that produce clustered embeddings and uses the discovered pattern as a detection signal even without access to labels or downstream heads. Meta-detection approaches (e.g., MNTD \cite{xu2021detecting}) train a higher-level detector over model responses or feature statistics across many simulated poisoning scenarios, enabling generalization to unseen attack types at the cost of requiring shadow-model training. 

For encoder-focused mitigation, recent proposals combine auditing with repair: mutual-information guided mitigation and related encoder-purification methods aim to isolate and remove backdoor-corrupted feature components in the encoder before downstream fine-tuning \cite{han2025mutual}. Finally, empirical studies (e.g., investigations into backdoored pre-trained models and BadEncoder-style attacks) show that encoder backdoors can transfer to many downstream tasks, underscoring the importance of specialized encoder scanning rather than relying only on downstream classifier checks \cite{shen2021backdoor,pan2023asset}. Feature-space auditing and encoder scanning provide powerful, targeted defenses for modern pipelines, but they also face limitations: they typically require white-box access to embeddings or the encoder, can be sensitive to choice of embedding layer and distance metric, and may be evaded by highly distributed or adaptive poisons that deliberately mimic clean embedding statistics; therefore they are most effective when combined with upstream data sanitization and downstream model inspection in a defense-in-depth strategy.

\subsubsection{Layerwise and Distributed Detection}  
Layerwise and distributed detection methods analyze representation anomalies ``across multiple layers'' (or jointly across layer pairs) to detect backdoors whose footprint is spread out through the network rather than concentrated in a single activation or neuron. The motivation for this family is simple: adaptive or transformation/frequency-based triggers often diffuse their signal so that no single hidden layer shows a strong, separable signature, yet the joint behaviour of several layers reveals subtle, consistent deviations from clean models. Practical techniques include (1) layerwise spectral or PCA scans that search for anomalous principal directions at several depths and combine those signals to amplify weak footprints \cite{tran2018spectral}, (2) robust multi-layer covariance/whitening and joint-clustering schemes that use robust statistics to highlight small but consistent shifts across layers (an approach related to SPECTRE \cite{hayase2021spectre}), (3) active neuron-stimulation carried out layer-by-layer to find distributed trojan responses and to reconstruct triggers from combined unit responses \cite{liu2019abs}, and (4) encoder/representation scanners that probe multiple intermediate embeddings (as DECREE does for pre-trained encoders) to detect backdoors that only become apparent when viewing several feature levels together \cite{feng2023detecting}.

Layerwise detection also benefits frequency and transform-aware analyses: frequency-domain trojans and warping-based attacks often produce complementary signals at different depths, so combining spectral inspections across layers increases detection power against Frequency based backdoors and WaNet-style constructions \cite{zeng2021rethinking,wang2021backdoor,nguyen2021wanet}. In addition, joint-layer nearest-neighbour and prototype consistency checks (extending Deep k-NN \cite{peri2020deep} ideas) can flag samples whose cross-layer embedding trajectories are inconsistent with their claimed label, catching poisons that evade single-layer k-NN tests. Finally, layerwise detection naturally guides repair: by identifying the layers that contribute most to the distributed signature, targeted pruning, fine-tuning, or layer-focused purification can be applied to remove or attenuate the backdoor while minimizing impact on clean performance. The main limitations are computational cost (multi-layer scans and joint statistics are heavier than single-layer tests), sensitivity to the chosen layers and metrics, and the fact that truly stealthy attackers can attempt to decorrelate signals across layers; nevertheless, layerwise and distributed approaches are currently among the most promising directions for detecting modern, highly-distributed backdoors.

\subsubsection{Parameter-space Anomaly Detection and Integrity Checks}  
Parameter-space defenses inspect a model's weights and parameter statistics (or protect their provenance) to detect tampering or implanted backdoors that are not readily apparent from inputs or outputs. One class of methods performs \emph{weight-space anomaly detection} by looking for unusual layer-wise statistics, low-rank perturbations, or targeted parameter shifts that correlate with malicious behaviour; such analyses are motivated by demonstrations that targeted weight perturbations can implant misclassification behaviours with minimal change to overall performance \cite{dumford2020backdooring,shen2021backdoor}. Complementing statistical scans, neuron-focused diagnostics (e.g., ABS \cite{liu2019abs}) actively probe internal units to reveal neurons or subnetworks that respond preferentially to triggers and whose connected weights therefore merit inspection or pruning; targeted pruning and fine-tuning (Fine-Pruning \cite{Liu2018FinePruningDA}) exploit this observation to remove or neutralize neurons that carry the backdoor while preserving clean accuracy. 

Black-box reverse-engineering tools (e.g., TABOR \cite{guo2019tabor}, DeepInspect \cite{chen2019deepinspect}) can also be used to generate candidate triggers whose existence implies suspicious parameter configurations and can guide parameter-space repair even without full white-box access. A second line of defense seeks to protect model integrity proactively via \emph{provenance, signing and watermarking}: cryptographic signing, reproducible build records, and model provenance logs (model ``DNA'') allow consumers to verify that a checkpoint originates from a trusted build and has not been tampered with in transit, while watermarking/fingerprinting methods (e.g., DeepSigns, Uchida-style watermarks) provide owner-level markers that help detect unauthorized model modifications or claim provenance in forensics \cite{rouhani2019deepsigns,uchida2017embedding,mu2023model}.

Combining provenance with parameter-space analytics gives a stronger guarantee: signing and attestation prevent casual supply-chain replacement, and weight-space scans or watermark checks raise alarms if signed artifacts have deviated from expected fingerprints.  
These parameter-space approaches are powerful in white-box or supply-chain settings but have practical limitations: weight-distribution scans can be evaded by ultra-fine or distributed perturbations that leave global statistics unchanged; neuron-stimulation and pruning require careful calibration to avoid harming clean utility; provenance and watermarking require secure operational practices and do not, on their own, detect backdoors inserted by an authorized builder; and many parameter-space checks are infeasible for third-party black-box model consumers. Therefore, parameter-space anomaly detection and integrity controls are best employed as part of a layered defense that includes dataset sanitization, representation auditing, and runtime monitoring.

\subsection{Target Tasks}
\label{sec:defenses-target-tasks}

Defenses often need to be tailored to the specific characteristics of the target computer-vision task because the attack surface, model architecture, and acceptable utility trade-offs vary substantially between tasks. Classification models operate on whole-image labels and typically use relatively compact output heads, which makes certain detection and repair strategies (e.g., per-class reverse-engineering, per-class activation clustering, and dataset-level sanitization) especially effective. In contrast, dense-prediction tasks (detection, segmentation) and temporal/video tasks require spatially or temporally-aware defenses. Consequently, it is useful to first review defenses that were designed for or primarily evaluated on image classification, both because they form the bulk of the literature and because many methods are later adapted to other CV (Computer Vision) tasks.

\subsubsection{Classification-centric Defenses}  
For image classification the defense literature is rich and well-studied; methods span the full pipeline: data sanitization, training-time hardening, post-training inspection/repair, and runtime input checks. Representation-based sanitizers such as spectral signatures \cite{tran2018spectral} and activation clustering \cite{chen2018detecting} detect and remove poisoned training samples by identifying separable structures in deep features, and robust refinements like SPECTRE \cite{hayase2021spectre} extend these ideas to weaker signals via robust covariance estimation. Post-training reverse-engineering (Neural Cleanse \cite{wang2019neural}) searches for minimal per-label triggers and flags anomalously small trigger solutions as evidence of a trojan; the recovered triggers are then used for mitigation (e.g., masking/pruning). 

Complementary runtime detectors such as STRIP \cite{gao2019strip} use randomized input perturbations and prediction-entropy tests to identify inputs whose outputs are invariant under heavy perturbation which is a hallmark of input-agnostic backdoors. Repair methods targeted at classifiers include neuron-level pruning and fine-tuning (Fine-Pruning \cite{Liu2018FinePruningDA}) which remove units disproportionately responsible for triggered behaviour, and black-box mitigation tools (DeepInspect \cite{chen2019deepinspect}, TABOR \cite{guo2019tabor}) that synthesize candidate triggers to guide repair when white-box access is limited. Training-time defenses adapted for classification, such as Anti-Backdoor Learning (ABL) \cite{li2021anti}, differential-privacy based updates (DP-SGD studies) \cite{10.1007/978-3-031-65172-4_20}, and adaptive sample-splitting (ASD) \cite{gao2023backdoor} aim to prevent backdoor memorization during optimization by isolating or down-weighting likely poisoned examples or by bounding individual sample influence. 

Detection using prototype / neighbour statistics (Deep k-NN \cite{peri2020deep}) and model-level litmus tests (Universal Litmus Patterns \cite{kolouri2020universal}, MNTD \cite{xu2021detecting}) provide alternative classification-oriented signals that generalize across diverse poisons by exploiting inconsistencies in embedding neighbourhoods or learned model behavior. More recent classification defenses include purification (e.g., Februus \cite{doan2020februus} and diffusion-based mitigation strategies \cite{zhou2024dataelixir}) that surgically remove suspected trigger regions or denoise invisible perturbations prior to classification.

Despite the breadth of methods, classifier-centric defenses share common limitations: many rely on the existence of a coherent activation or spectral signature and so can be bypassed by sample-specific or carefully distributed triggers; runtime detectors trade off latency and false positives; reverse-engineering can be computationally expensive and fragile for complex triggers; and training-time interventions require control of or access to the training pipeline. Importantly, techniques that work well on standard classification benchmarks (CIFAR/ImageNet) do not automatically translate to detection or segmentation tasks without task-specific adaptation. For these reasons, robust classification defense practice typically combines several orthogonal techniques (sanitization + training hardening + post-training inspection + runtime checks) and evaluates them under adaptive (defense-aware) attacks to measure practical resilience.

\subsubsection{Object Detection-specific Defenses}  
Defenses for object detectors adapt and extend classification-oriented techniques while adding detector-aware mechanisms such as proposal-level sanitization, spatial-consistency checks, and object-proposal auditing. A common strategy is to apply dataset sanitization and representation-level filtering at the image and object-proposal level: spectral and activation-based filters (e.g., spectral signatures \cite{tran2018spectral}, activation clustering \cite{chen2018detecting}, SPECTRE \cite{hayase2021spectre}) can be used on per-proposal embeddings to flag suspicious training instances that produce anomalous proposal-level representations. Post-training, reverse-engineering and trigger-synthesis techniques (Neural Cleanse \cite{wang2019neural}, DeepInspect \cite{chen2019deepinspect} and TABOR \cite{guo2019tabor}) have been adapted to detectors by searching for minimal perturbations or local patches that systematically alter object predictions across proposals; when found, these patterns guide masking, inpainting, or targeted pruning of the model.

Runtime defenses for object detection typically examine consistency of predictions under perturbation. For example, STRIP-style methods randomly perturb the input (e.g., adding noise, overlays, or corruptions) and then test whether the detector’s predictions change. In a benign model, such perturbations usually alter bounding boxes or object scores. However, in a backdoored detector, trigger-activated inputs often produce invariant predictions, i.e., the same object detections persist regardless of the perturbation. This abnormal stability is a signature of input-agnostic backdoors and can be revealed by analyzing proposal crops or detector feature maps under such perturbations \cite{gao2019strip,liu2023detecting}.

Input-purification (e.g., Februus \cite{doan2020februus}) and proposal-level inpainting have also proven effective for patch-style attacks in object detectors by removing candidate trigger regions before detection. For physical-world and location-flexible attacks, spatial-aggregation and multi-view consistency checks (comparing detections across scales, augmentations or temporal frames) help detect triggers that only manifest under certain views or positions \cite{zhao2020clean}. Model-repair techniques such as fine-tuning on a small trusted object-level dataset \cite{Liu2018FinePruningDA}, selective pruning of proposal-head neurons \cite{shen2021backdoor}, and targeted layerwise fine-pruning \cite{li2023badlidet} have been shown to remove or reduce detector backdoors while preserving localization performance when applied carefully.

Finally, object detection-specific defenses must be evaluated under multi-instance metrics (e.g., mAP changes, false negative/positive per-class) and in realistic scenarios (physical triggers, occlusion, varied lighting); several recent surveys and empirical studies document that many classification defenses degrade or fail when naively applied to detectors, motivating detector-tailored methods and benchmarks \cite{gao2020backdoor}. In practice, robust protection for object detection combines proposal-level sanitization, runtime proposal-consistency tests, and cautious model repair, and it requires attacker-aware evaluation using detector-specific attacks (e.g., cloaking, targeted disappearance, and trigger-position-flexible attacks) to measure real-world resilience.

\section{Conclusion}
\label{sec:conclusion}

Backdoor attacks constitute a diverse and evolving threat to computer-vision systems. The literature spans a wide spectrum of techniques: from straightforward patch-stamping and dirty-label poisons to sophisticated clean-label, input-aware, frequency-based, and parameter-space attacks that are resilient to fine-tuning and compression. Complementary defenses that include: dataset sanitization, robust training, post-hoc inspection and repair, and runtime detection/purification have made meaningful progress, but no single class of defenses yet offers universal protection. In particular, methods tuned to detect reusable, localized triggers often fail against sample-specific, distributed, or parameter-space backdoors, and many reported defenses are insufficiently evaluated against adaptive attackers.

\subsection{Practical recommendations}
Based on the surveyed works, we recommend the following practices for practitioners and evaluators:
\begin{enumerate}
  \item \textbf{Layered defenses:} combine data-centric measures (sanitization, robust augmentation), model-inspection techniques (activation and weight auditing), and runtime integrity checks (provenance, signed checkpoints). Supply-chain threats and parameter tampering typically require platform-level integrity controls in addition to algorithmic defenses.
  \item \textbf{Adaptive evaluation:} evaluate defenses under adaptive attackers that are aware of the defense mechanism (e.g., craft sample-specific triggers or parameter-space attacks designed to evade the proposed method).
  \item \textbf{Cross-task benchmarks:} benchmark attacks and defenses across classification, detection and segmentation tasks to avoid overfitting to narrow datasets (e.g., CIFAR/ImageNet-style settings).
  \item \textbf{Report comprehensive metrics:} publish both clean utility (CA, mAP, IoU) and attack metrics (ASR) across multiple trigger strengths, poisoning rates, and physical transformations. Include runtime and false-positive characteristics for detectors.
\end{enumerate}

\subsection{Open directions}
Important research directions that remain insufficiently explored include:
\begin{itemize}
  \item \textbf{Parameter-space and supply-chain attacks:} more realistic threat models, detection tools for tampered checkpoints, and cryptographic provenance mechanisms.
  \item \textbf{Certifiable defenses:} tractable certification methods for restricted trigger families (for instance, localized patches) and better theory connecting representation structure to backdoor vulnerability.
  \item \textbf{Physical robustness and transferability:} study of how backdoors persist under real-world imaging conditions and when pre-trained encoders are transferred to downstream tasks.
  \item \textbf{Federated and edge settings:} practical defenses for federated model-poisoning and constrained-edge devices that are vulnerable to hardware fault attacks (e.g., targeted bit-flips).
\end{itemize}

\bibliography{ref}

\end{document}